# FLUHOLOSCOPY – COMPACT AND SIMPLE PLATFORM COMBINING FLUORESCENCE AND HOLOGRAPHIC MICROSCOPY

**David Alonso, Javier Garcia and Vicente Micó\***

Departamento de Óptica y de Optometría y Ciencias de la Visión, Facultad de Física, Universidad de Valencia, C/Doctor Moliner 50, Burjassot 46100, Spain
\* Correspondence: vicente.mico@uv.es

**Abstract:** The combination of different imaging modalities into single imaging platforms has a strong potential in biomedical sciences since it permits the analysis of complementary properties of the target sample. Here, we report on an extremely simple, cost-effective, and compact microscope platform for achieving simultaneous fluorescence and quantitative phase imaging modes with the capability of working in a single snapshot. It is based on the use of a single illumination wavelength to both excite the sample's fluorescence and provide coherent illumination for phase imaging. After passing the microscope layout, the two imaging paths are separated by using a bandpass filter and the two imaging modes are simultaneously obtained by using two digital cameras. We first present calibration and analysis of both fluorescence and phase imaging modalities working independently and, later on, experimental validation for the proposed common-path dual-mode imaging platform considering static (resolution test targets, fluorescent micro-beads and water-suspended lab-made cultures) as well as dynamic (flowing fluorescent beads, human sperm cells and live specimens from lab-made cultures) samples.

## 1. Introduction

Multimodal imaging deals with mixing of different imaging modalities in a single imaging platform in order to gain deeper knowledge about a specific sample [1]. Some examples of such integration of imaging modalities can be found, just as examples, in biology [2], medicine [3] and ophthalmology [4] as well as the combination with deep learning approaches [5,6]. In the frame of this multimodal strategy, optical microscopy is one of the most appealing fields from a multimodal imaging perspective where combination of non-linear [7–9], linear [10–13] or a mixing [14–16] of techniques allows parallel and complementary information of a specific sample or event to improve imaging and understanding of specific biological processes.

In particular, strong efforts have been provided in the last two decades for combining fluorescence imaging (FI) with quantitative phase imaging (QPI) [17–29]. This dual-mode imaging modality combination is probably the simplest one capable of working in a single exposure while maximizing the obtained information. While QPI retrieves morphological information and cell dynamics coming from the topography and/or spatial refractive index distribution in the specimen [30], FI can reveal functional features for studying sub-cellular molecular events of live cells by using fluorophore labeling with high specificity [31]. For this reason, the combination of FI with QPI opens up new possibilities to extract more information about biological specimens and processes.

However, the integration of FI and QPI into a single imaging platform sometimes yields into some drawbacks. In some cases, such a combination is implemented using a single digital camera for recording the images [17,18,23,25,26]. Thus, one needs to switch between the two illumination modes and their associated optical paths meaning that there is no possibility for single snapshot imaging and the recordings must be done sequentially in time preventing the imaging of some biological events that occur at very short time scale This disadvantage can be circumvented by using dynamic range multiplexing where both images are recorded in a single snapshot at the expenses of reducing the dynamic range of the detector [20,24]. Nevertheless, the preferred solution implies the use of 2 different synchronized digital cameras (one for QPI and the other for FI modes) where a dichroic mirror separates both imaging paths [12,20–22,27–29].

Another drawback comes from the fact that both imaging modes utilize different geometries (FI is usually implemented in reflection while QPI in transmission configurations) [17,19,22–24,26–29]. As a result, the whole imaging platform becomes bulky including some additional optical components (dichroic mirrors, selective filters, etc.) that increase system's price and weight. This can be a limiting issue since compact and cost-effective systems could be pursued depending on the application such as, for instance, the study of a specimen inside an incubator or in the field-setting as well as to adapt the instrument to resource limited labs. This second drawback can be partially reduced by introducing the fluorescence excitation beam using an oblique angle geometry in transmission with an illumination angle higher than the one defined by the numerical aperture (NA) of the used objective lens [25]. This way the fluorescence excitation beam will not pass through the imaging path and some optical components can be spared and the system is reduced in size.

In this manuscript we propose a simple, compact and cost-effective imaging platform for the integration of FI and QPI into a common-path system with the capability of working in a single snapshot (both imaging modes simultaneously in parallel). The key point is to use a single illumination light with a double function: to excite the fluorescence in the sample and to provide the holographic imaging. Thus, the system is minimized in terms of optical components improving compactness, weight and pricing. After the tube lens, a regular dichroic mirror separates both imaging modes and 2 digital cameras allow simultaneous recordings of the same specimen under test. FI is obtained in the whole field of view (FOV) provided by the microscope lens and limited by the used fluorophore while QPI comes from an in-line configuration similar to the configuration on phase from defocus images [25,26] and phase imaging under Gabor regime [32]. Experimental results are included considering different types of samples, that is, static (resolution test targets, fluorescent beads and lab-made cultures in water suspension) and dynamic (flowing fluorescent beads, human sperm cells and live specimens from lab-made cultures) samples. Summarizing, the proposed dual-mode imaging platform provides an optimized configuration for FI/QPI integration that minimizes size, weight and price (thus improving compactness, portability and affordability) and that can be easily configured by selecting some specific components (coherent illumination, fluorophore and dichroic mirror) depending on the target application. Through this manuscript, Section 2 provides system's description, preparation of samples and different details concerning the imaging procedure, Section 3 presents calibration of QPI and FI modes as well as experimental support and proof of principle validation of the integrated platform. Finally, Section 4 discusses and concludes the paper.

## 2. Materials and Methods
### 2.1. Experimental setup and numeric propagation

The proposed setup must allow the simultaneous capture of the incoherent light emitted by fluorescence and the coherent in-line Gabor hologram of the sample. To achieve this goal, we put forward a simple system (Fig. 1) composed of a Mitutoyo long working distance objective (10X 0.28NA) with coherent illumination coming from a fiber coupled diode laser. Specifically, we have used blue laser light (450 nm) of the multiwavelength illumination source from Blue Sky Research (SpectraTec 4 STEC4 405/450/532/635 nm). Before illuminating the sample, light is collimated using a collimation lens (QIOptics

G052010000, achromat visible, 40 mm focal length) module which is mounted along a vertical axis to allow horizontal sample positioning. The sample is held by a translation stage which allows the vertical displacement of the object by an adjustment knob. Light is deflected by a mirror (Thorlabs BB111-E02, broadband dielectric mirror) after passing the microscope objective to a horizontal plane where the rest of the system lays. After being reflected, both fluorescent and coherent beam paths are focused using a tube lens system(Thorlabs TTL180-A, 180 mm focal length) and separated with a long pass filter dichroic mirror (Thorlabs DMLP490, Long pass Dichroic Mirror, 490 nm Cut-On). Thus, hologram and direct fluorescent imaging of the sample can be simultaneously captured using two analogous digital cameras (Basler acA1300, 1280x1024 px, 4.6 µm pixel size) connected to the visualization software. To easy understand the optical layout, the coherent beam path has been included in 3D perspective at Fig. 1 to clearly identify its positioning along the horizontal plane.

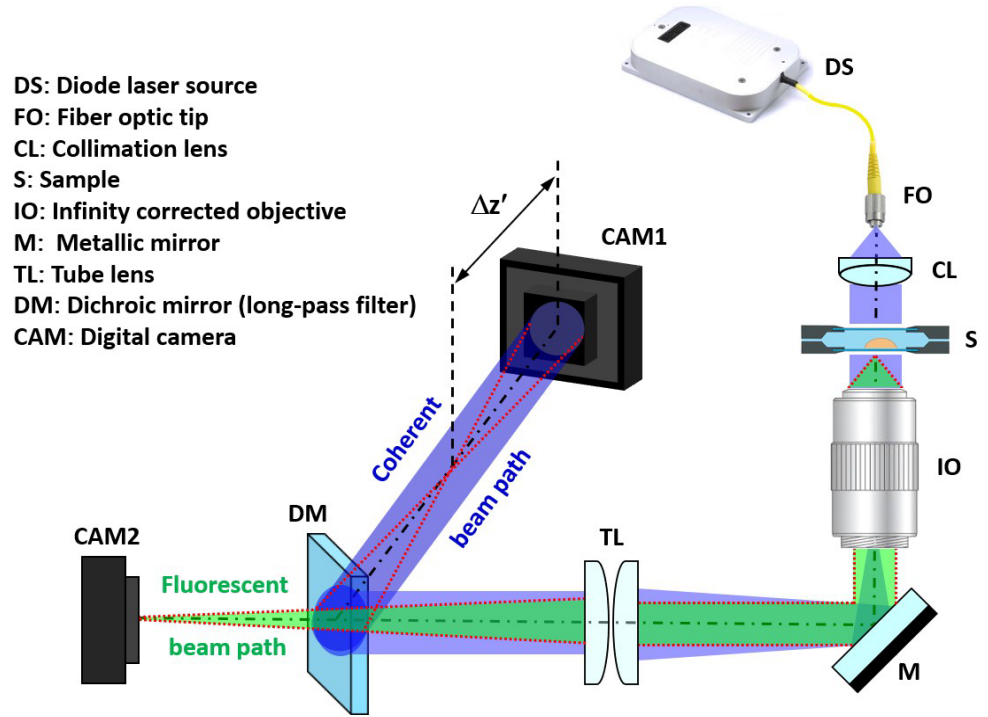

**Figure 1.** Optical arrangement of the proposed system. Blue beam represents the coherent laser light, while the green one is the fluorescent emission. Red dashed line represents the ray tracing of the image at the output plane where the camera for FI (CAM2) is placed and from where $\Delta z'$ is generated according to the optimal defocus distance necessary for a proper phase reconstruction.

Another key aspect of the assembly is that in order to induce a Gabor hologram recording scheme, the camera (CAM1) must be axially displaced from the image plane. This defocus distance is not arbitrary and has to be calibrated in order to obtain good results (will be discussed in section 3.1.1). Once the hologram is recorded, it is digitally propagated by the use of the convolution method applied to the diffraction Rayleigh-Sommerfeld integral [33]. Here, the diffraction integral is numerically computed by using three digital Fourier transformations as:

$$U_0(x, y; d) = FT^{-1}\{FT\{U(x,y)\} \times FT\{h(x, y; d)\}\} \qquad (1)$$

where $U_0(x,y;d)$ is the numerically propagated complex amplitude wave field at the object plane, FT is the numerical Fourier transform operation ( implemented by fast Fourier transform (FFT) algorithm, $U(x,y)$ is the amplitude at the recording plane coming from the recorded intensity distribution (in-line Gabor hologram), $h(x,y;d)$ is the impulse response of free space propagation, (x, y) are the spatial coordinates and d is the propagation distance. Equation 1 can be simplified by defining the Fourier transformation of the impulse response as $H(u, v; d) = FT\{h(x, y; d)\}$. Thus, the calculation of the propagated wave field is simplified to

$$U_0(x, y; d) = FT^{-1}\left\{\hat{U}(u, v) \cdot H(u, v; d)\right\} \quad (2)$$

*2.2. Calibration targets and imaging of the samples*

For the calibration of both imaging modes (FI and QPI) we have also used amplitude and phase USAF test targets. In particular, in phase imaging calibration we used different USAF style phase resolution tests of the quantitative phase target from Benchmark Technologies (www.benchmarktech.com/quantitativephasemicroscop), having nominal heights of 50 μm , 100 μm and 150 μm. In the case of the FI calibration, an USAF positive resolution test target (amplitude test target) from Edmund Optics was used. In order to produce the fluorescence emission, a drop of fluorescein was deposited on top of it and subsequently covered with a coverslip.

Imaging of the samples after positioning the CAM1 at the optimal defocus distance consist on focusing the FI mode in CAM2 by adjusting the vertical adjustment knob of the micrometric translation stage. In this way, the hologram will be captured by CAM1 having the optimal defocus for a properly phase reconstruction. Regarding the process of acquisition, since the light emitted during the fluorescence process is much dimmer than the one coming from the coherent imaging path, it will be needed a long exposure time to get enough intensity images. Specifically, all the images corresponding to the FI mode (CAM2) have been acquired during 100 ms integration time while the holographic ones (CAM1) are obtained using a much shorter integration time (around 0.1 ms but depending on the laser intensity). However, to synchronize both imaging modes (same snapshots at same time) while do not produce huge data in the holographic recordings, the acquisition framerate has been fixed to the more restrictive one, that is, 10 fps coming from CAM2.

*2.3. Preparation of fluorescence micro-beads and living samples.*

Micro-beads are an excellent sample for calibration of both cameras owing to the simplicity of working with them and the well-defined shape and fluorescence properties. We have used a 10 microns diameter fluorescent microsphere suspension (Thermo scientific G1000) with a maximum excitation/emission wavelengths of 468/508 nm, respectively, so they can be excited with the 450 nm laser light and will emit above 490 nm which is the cutoff wavelength of the used bandpass filter. These beads are distributed in a liquid solution, but the density is too high for phase reconstruction using Gabor holography (which requires a weak diffracting object). Hence, we prepared a wide range of samples adding different quantities of distilled water in order to obtain a properly beads density. Subsequently, these samples are introduced by pipetting and sealed in 20 μm thickness counting chambers in order to prevent aqueous medium evaporation.

On the other hand, human sperm cells and water suspended microorganism have been used as living samples for testing our proposed imaging platform. In particular, lab-made cultures have grown at room temperature in a container filled with water and some plant residues. To produce fluorescent emission, a general eye drop solution fluorophore for ocular use (minims fluorescein sodium 2% provided by BauschLomb) have been added in every test. However, since the used fluorophore usually works combined with the eye tear, it has been diluted in a water solution of 1:10 before spreading it on the samples. Next, biosamples are dropped into an object holder by pipetting 10 to 15 milliliters from the eppendorf where samples are prepared (diluted or directly extracted from the main culture solution) and are sprinkled with one load of the water solution with fluorescein. In some cases, the samples are prepared in conventional microscope slides so they are covered with a coverslip before placing them in the sample stage.

## 3. Results

*3.1. Phase imaging calibration*

### 3.1.1. Defocus distance

As we mention before, one important step to achieve high quality phase images is to select an appropriate defocus distance ($\Delta z'$). Since precisely varying the position of the camera in a microscope embodiment is a complicated task, we have decided instead to vertically shift the object by a certain amount $\Delta z$. Note that both distances can be related through the square of the objective magnification (M) as $\Delta z' = -M^2 \Delta z$. However, obtaining the precise defocus distance experimentally is not a trivial procedure because it depends on the specific objective characteristics (beam divergence and distances hard to measure with accuracy). Hence, we have performed the following analysis using the 150 nm - thickness USAF-style phase resolution test: we have varied z in the range 300 μm < $\Delta z$ < 1800 μm using steps of 50 μm and compute numerically the best in-focus image for each distance. A sweep has previously been done to check qualitatively which is the best resolution zone. The experimental methodology for the phase target consist on placing the target in the microscope stage and focus on the sample plane by vertically moving the test using the coarse/fine adjustment knob of the micrometric translation stage. This defines the regular imaging condition in the microscope layout and would be our starting point ($\Delta z = 0$) from which defocus distances are produced using the vertical micrometric adjustment of the microscope's sample stage. Note that we have not considered negative defocus distances due to the worse performance in phase reconstruction [32].

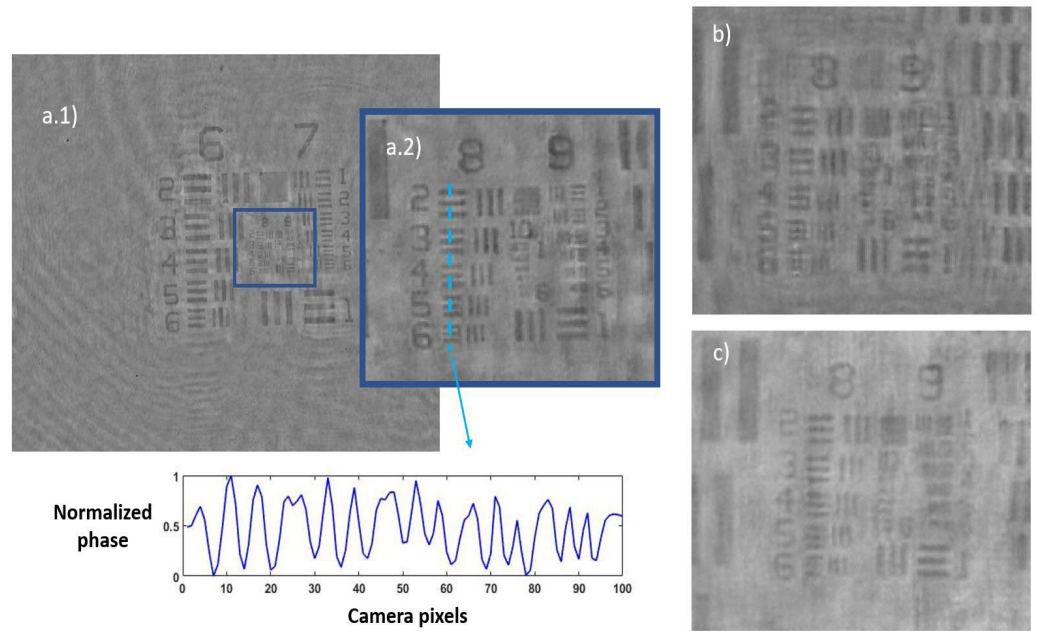

**Figure 2.** Phase reconstruction images of the USAF test for three different defocus distances: **(a.1)** presents the obtained image from best defocus distance while **(a.2)** show a noise-filtered magnification of the central area with the finest details of the test. The blue dashed line represent the zone where the contrast parameter has been compute, showing the horizontally averaged and normalized phase value across the group at the bottom picture. On the other hand **(b)** and **(c)** present magnifications of the central area of the USAF test for a defocus distance of 300 μm and 1800 μm respectively, showing worse reconstructed images.

Once the set of different reconstructed phase USAF target images from each defocus distance is obtained, we can easily get the effective resolution as the last resolved element for each in-focus image. Equation 2 represents how to numerically propagate the recorded hologram to focus the sample plane. We have implemented this algorithm in Matlab software by varying the propagation distance (d in Eq. 2) and looking the visual quality of the retrieved image. For this demonstration, we have not included any technique (iterative algorithm for twin image mitigation, averaging for noise reduction, etc) for improving the reconstructions. Moreover, the sample is simple focused by visual criterion looking at the reconstructed images even that some automation can be implemented in the reconstruction process. Thus, three cases are exemplified in Fig. 2 for best defocus distance (around 1100 µm), 300 µm and 1800 µm. Figure 3 shows the results of this analysis by plotting resolution vs. defocus distance. We can see an optimal range for Δz between 700 and 1450 µm where the resolution is maximal (group 9, element 2 defining a resolution limit of 1.74 µm). If we are above or below this defocus range, resolution starts to get worse (group 9 is not resolved), as we can see in Fig. 2b and 2c.

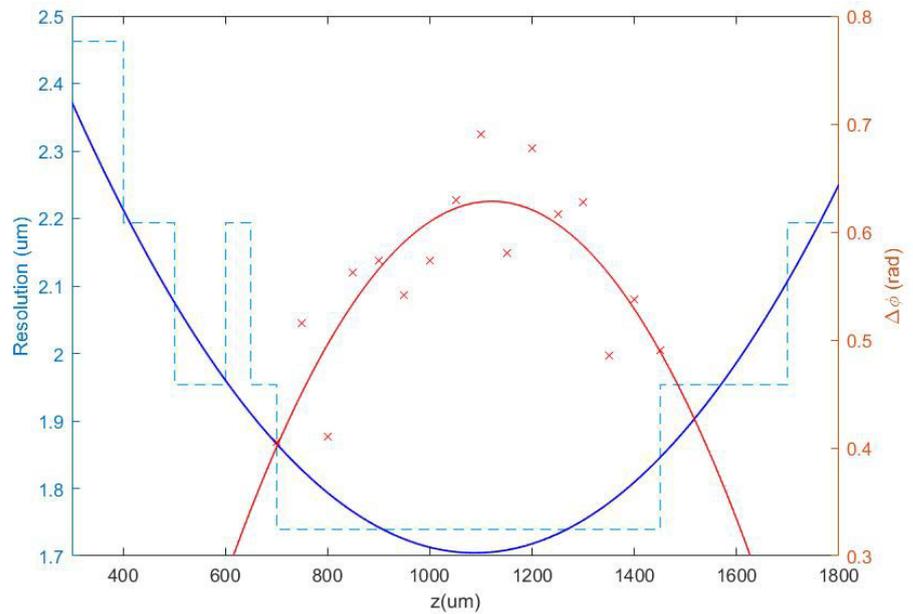

**Figure 3.** Representation of the resolution obtained from the analysis of the phase test images for different defocus distances in the range 300 µm < Δ z < 1800 µm (blue dashed line), as well as a contrast parameter calculated on the range where the resolution limit is achieved (red crosses). Continuous lines show quadratic polynomial fits to the respective data.

At this point, we could choose a defocus distance around 1100 µm in the center of the flat region of best resolution limit (minimum value of the blue plot at Fig. 3). Notice as there is some tolerance around this value from which we will not lose resolution due to small mismatches in the defocus distance of the camera. In order to obtain a more quantitative result and to support the best defocus distance driven by the resolution criterion analysis, it is possible to include another parameter in the picture: the phase amplitude value defined as Δϕ = *mean*(*maximum_peaks*) − *mean*(*minimum_peaks*). Thus defined, Δϕ is a kind of contrast value coming from the phase step produced by the test target which is supposed to be maximum at best defocus distance. We have computed it for the group 8 of the best reconstructed in-focus images for each defocus distance. Through the red plot at Fig. 3 we can see the computed contrast values depending on the defocus distances. Despite the fact that the dispersion of points is quite large, they show a behavior similar to the resolution criterion analysis, being maximum around the same defocus value. For this reason, we   212 have chosen 1100 μm as proper defocus distance for holographic phase reconstruction in this setup, which results in a camera shift of $\Delta z' = 11$ cm at the image space.

### 3.1.2. Quantitative validation

Once the best defocus distance has been selected, quantitative validation with experimental results is presented. For this purpose, we have used the three different heights mentioned of the USAF phase target to perform quantitative phase imaging (QPI) analysis. Figure 4a shows how the increase in the thickness of the elements suppose an increase in the optical phase. Furthermore, computing the phase step ($\Delta\phi$) between elements and background (Fig. 4b) allows obtaining the height of the phase target ($\Delta t$) through equation:

$$\Delta t = \frac{\lambda \Delta \phi}{2\pi(n_{test} - n_{medium})} \quad (3)$$

where $\lambda$ is the illumination wavelength (450 nm in our case), while $n_{test}$ is the refractive index of the phase target material (Corning Eagle XG glass, $n_{test}$ = 1.5185 according to https://refractiveindex.info/) and $n_{medium} \approx 1$ as the refractive index of the air (in this particular case).

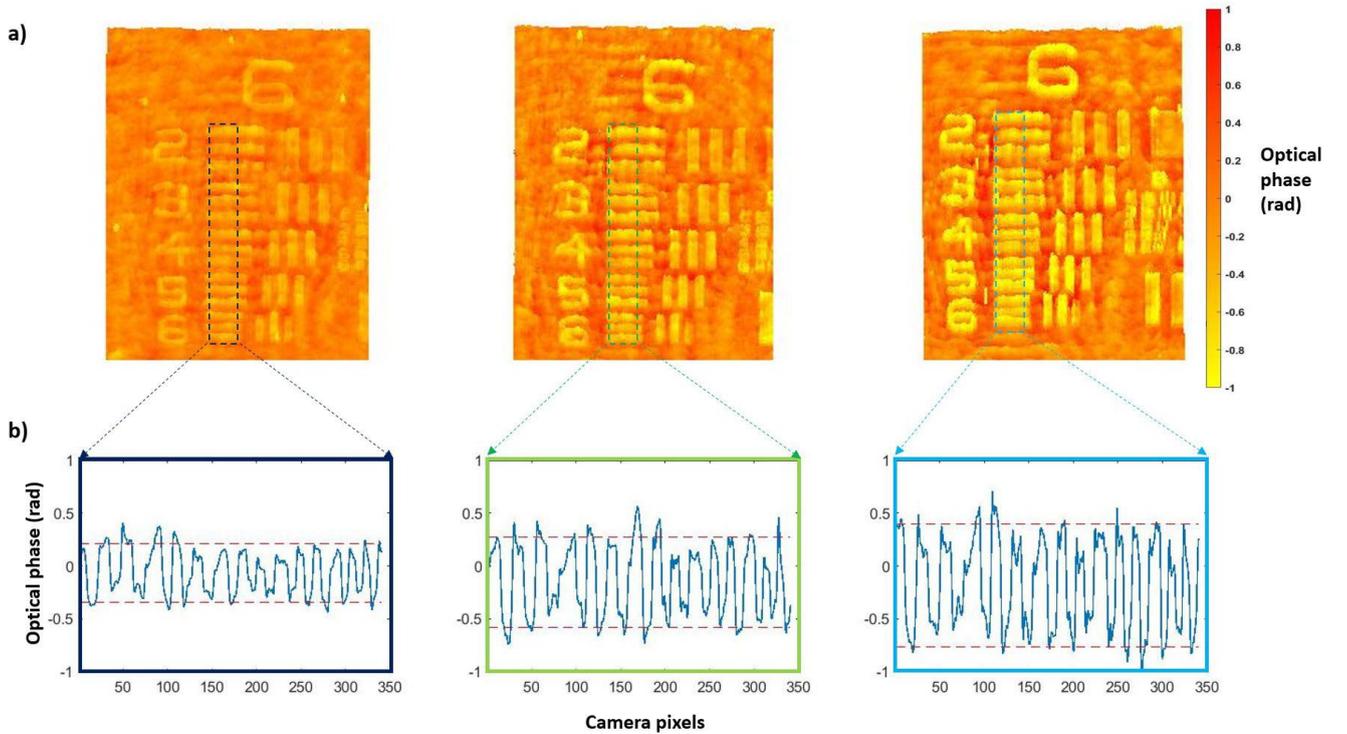

**Figure 4.** Height analysis performed on the group 6 of three different thicknesses (50, 100 and 150 nm nominal heights) for the USAF phase target. **(a)** Images of the retrieved phase distributions by numerical propagation of the different test targets. **(b)** Phase profile computed along the dashed marked regions corresponding with elements 2 to 6. Phase step values increase from left image (50 nm) to the right one (150 nm) and they are retrieved as the gap between red dashed lines at **(b)** that represents the average of the maximum and minimum peaks of retrieved optical phase.

Results of this analysis are shown in Table 1, where the height of the phase target obtained from Equation 3 is compared with the value provided by the manufacturer (not the nominal but the measured one). We can see a good agreement for the two thicker values (100 and 150 nm heights) where the expected measured value is inside the error provided by the presented quantitative phase characterization. However, the measured QPI value is higher in comparison with the one reported by the manufacturer for the lowest target height value (50 nm). We think this is due to the noise coming from the in-line geometry due to the presence of twin image disturbance and coherent artifacts that are present when implementing a Gabor in-line configuration. All these drawbacks are also responsible for the deviation between real and measured values in the other two characterized heights. However and despite all the shortcomings provided by the Gabor layout, we still think it is advantageous because it defines the most simple holographic layout possible as well as retrieves QPI values quite close to the real ones.

**Table 1.** Comparison between the thickness of the elements (Δ$t$) obtained from the phase step analysis (Δ$\phi$) and the values (nominal and measured ones) provided by the manufacturer.

| Nominal values (nm) | Measured values (nm) | Retrieved phase steps Δ$\phi$ (rad) | Retrieved thickness values Δ$t$ (nm) |
|---|---|---|---|
| 50 | 59.1 | 0.56 ± 0.11 | 77 ± 15 |
| 100 | 114.4 | 0.84 ± 0.17 | 120 ± 20 |
| 150 | 173.3 | 1.16 ± 0.16 | 160 ± 20 |

The holographic architecture has been extended to more complex samples. Specifically, 10 µm - diameter beads as well as water suspended microorganisms have been selected to validate QPI coming from the proposed holographic layout. These samples are processed according to Section 2.3. Figure 5a shows the recorded hologram of a low-density beads sample while Fig. 5b includes the retrieved phase distribution by numerically propagating to the in-focus beads plane.

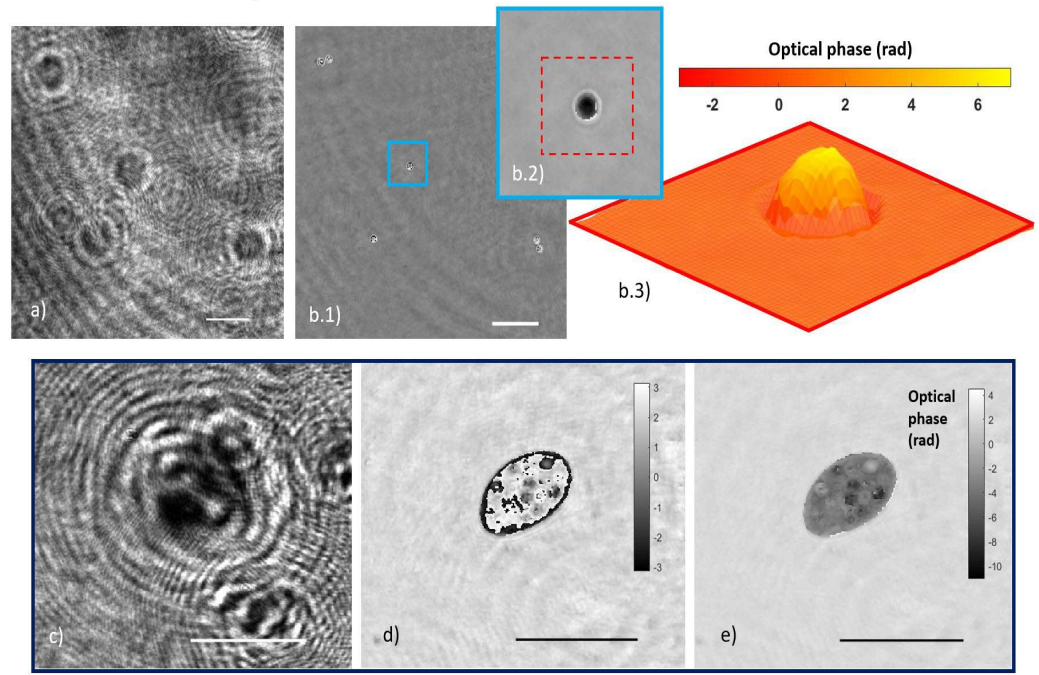

**Figure 5.** Holograms and phase distributions of different samples. **(a)** Hologram of a low-density water suspended beads sample. **(b.1)** Phase distribution retrieved by numerical propagation using the defocus distance obtained in calibration while **(b.2)** and **(b.3)** show a magnification of a single bead after applying an unwrapping procedure and being represented in 2 and 3 dimensions respectively. **(c)**, **(d)** and **(e)** Representative frames corresponding, respectively, with movies of the hologram, the reconstructed wrapped and unwrapped phase sequences of a water suspended microorganism (Visualizations 1, 2 and 3). Scale bars represent a length of 50 µm.

Since beads diameter is enough to produce a phase step greater than 2π, wrapping effects appear in the reconstructed phase images. Although it is possible to unwrap the phase in order to perform QPI, as figures 5(b.2) and 5(b.3) depict, phase reconstructed from static microbeads shows defects around the bead's edge. The main reasons of this problem are the low NA of the objective used and the high refractive index of the beads, causing that only a small cone of light to be captured by the objective lens thus physically preventing an accurate reconstruction. Hence, only a spherical cap of the bead can be reconstructed and the retrieved phase information becomes noisy with lack of information towards the bead's edge. This effect is less aggressive than in the case of air suspended beads (higher refractive index step), but still enough to cause the beads appearing surrounded by artifacts. On the other hand, Figs. 5c, 5d and 5e present a frame of the movies corresponding with the hologram (Visualization 1) and the reconstructed wrapped (Visualization 2) and unwrapped phase (Visualization 3) distributions, respectively, of a living organism in free movement on the aqueous medium where phase imaging allows to observe some structure inside the organism.

### 3.2. Fluorescence imaging calibration

In order to validate the performance of the FI mode, we have used the two different types of fluorophores previously mentioned: fluorescent beads and eye drops solution. Figure 6a shows the stained USAF positive resolution test target. That way, the image is observed in reverse contrast since it is the background itself the one emitting the fluorescent light, contrary to what happens in Fig. 6b where the fluorescent beads are the emitters themselves. Figure 6c shows a micro-organism moving on the aqueous medium developed at lab conditions (Visualization 4). The brighter areas are due to the differential fluorescein adherence to the different structures in the biosample (we can see some more brilliant parts corresponding with fluorescein accumulus at those parts).

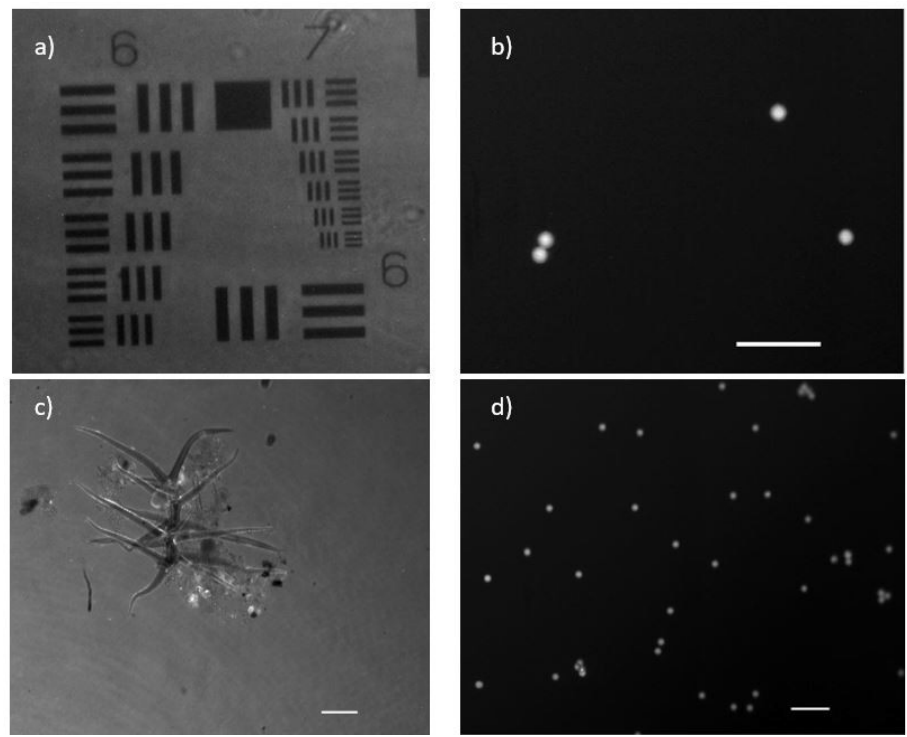

**Figure 6.** Images and visualizations of fluorescent samples acquired with an exposure time of 100 ms. **(a)** USAF amplitude test with added fluorescein. **(b)** Low density sample of fluorescent micro-beads (10 μm in diameter). **(c)** Stain environment with fluorescein showing a micro-organism in free movement (Visualization 4). **(d)** Mid-density sample of fluorescent micro-beads flowing on a aqueous medium (Visualization 5). White scale bars are 50 μm length.

On the other hand, through Fig. 6d, we can see the movement of a water suspended fluorescence beads flowing into the counting chamber (Visualization 5). Both movies show that the proposed frame rate (discussed in Section 2.2) is enough to track the sample's movement (at least considering the movement speeds of the objects here considered) as well as the total suppression of the coherent illumination which is not introducing veil effects. In addition, we have computed some quality imaging ratios to evaluate the performance of the FI mode. Thus, signal-to-noise ratio (SNR) and signal-to-background ratio (SBR) are calculated from Fig. 6b. This figure contains 4 fluorescent beads over dark background and it is nice example to compute those parameters in our system. We have applied the following definitions to SNR and SBR [34]:

SNR = (mean signal - mean background)/(STD background), and
SBR = mean signal / mean background.

Using those definitions, we have obtained the following results: SNR = 9.2 ± 0.9 and SBR = 2.95 ± 0.15 where the standard deviation error has also been included for every metric. All of them are good results since the SNR value approaches to 10 (the signal is one order of magnitude higher than the noise) while the SBR is higher than 2 (value from which sample can be clearly differentiated from background).

### 3.3. Combined station

After independent calibration and validation of our two introduced imaging modalities (QPI and FI), this section presents the results provided by the combined imaging platform where simultaneous dual mode imaging is achieved. As first case, we use static samples (micro beads inside the counting chamber filled with water and lab-made culture in water suspension). Figure 7 includes the simultaneously reconstructions coming from the two experimental imaging modes when considering the micro spheres. FI shows a perfect contrast visualization mode (bright beads with dark background) where only light from the beads is collected. And QPI shows additional information of the sample since we can observe the liquid edge (something fully missed in the FI mode) meaning that the micro bead solution is not completely covering the full FOV. The difference about the information provided by fluorescence and holography is also evident at Fig. 8 where the case of the lab-made culture in water suspension is presented. Here, the fluorescein adheres or accumulates to the area of the cell nuclei while phase imaging shows the entire cell spatial distribution. As in previous case, FI is bright against dark background while QPI is complementary (clear background with darken cells) to it.

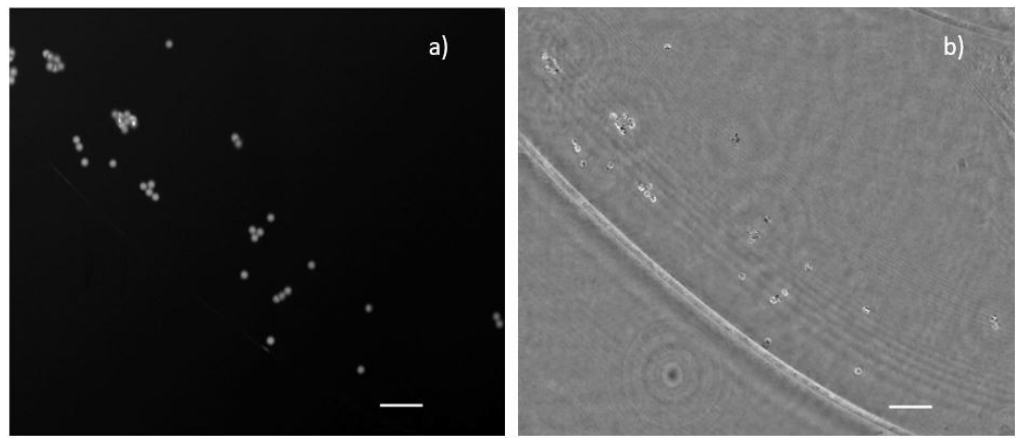

**Figure 7.** Simultaneous images of a mid-density water suspended fluorescent beads sample. **(a)** Direct image from fluorescent emission. **(b)** Phase distribution retrieved after numerical propagation to the in-focus plane. Scale bars are 50 μm length.

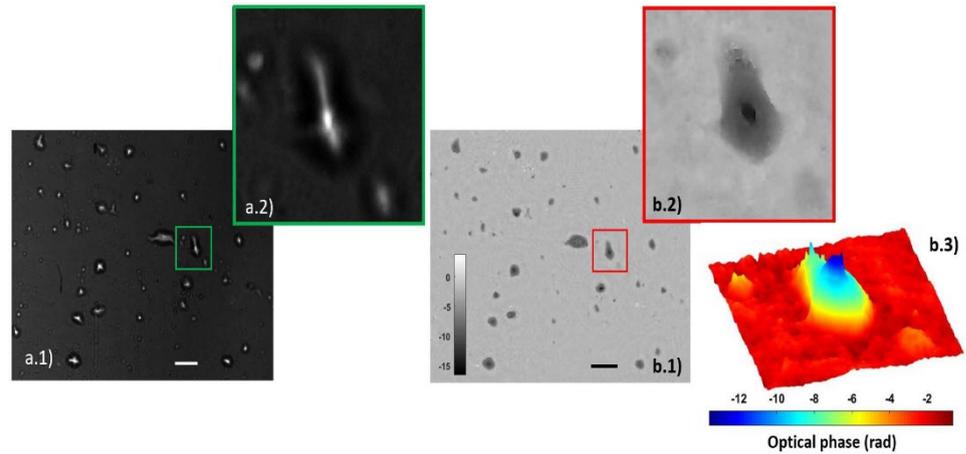

**Figure 8.** Lab-made cultures cells stained with fluorescein. **(a.1)** Full field image of the fluorescent arm while **(a.2)** includes a magnification of a single cell. **(b.1)** Full field image of the phase distribution while **(b.2)** and **(b.3)** depict the magnification of the same single cell in 2 and 3 dimensions respectively. Scale bars are 50 µm length.

As second case, living/moving samples are considered under the proposed visualization platform. As previously mentioned, movies have been acquired with a frame rate of 10 fps considering the case of beads, lab-made cultures and human sperm cells. Results are presented showing fluorescence and holography imaging as separated images (Fig. 9) for flowing beads or combining both in the same visualization frame where half FOV is for FI and the other half for QPI (Fig. 10) for living cells. Note as, in this latter case, some micro-organism and sperm cells are able to cross the separation line between the two imaging modes, allowing visualization with the two imaging modalities inside the same movie. While the lab-made culture has a number of microorganism large enough to be observed and perfectly resolved, the resolution of the system is in the limit to resolve the sperm cells because spermatozoa tail's width is below 1 µm and exceeds the resolution limit; thus, the retrieved phase distribution only shows clearly the head of sperm cells in free movement inside the counting chamber.

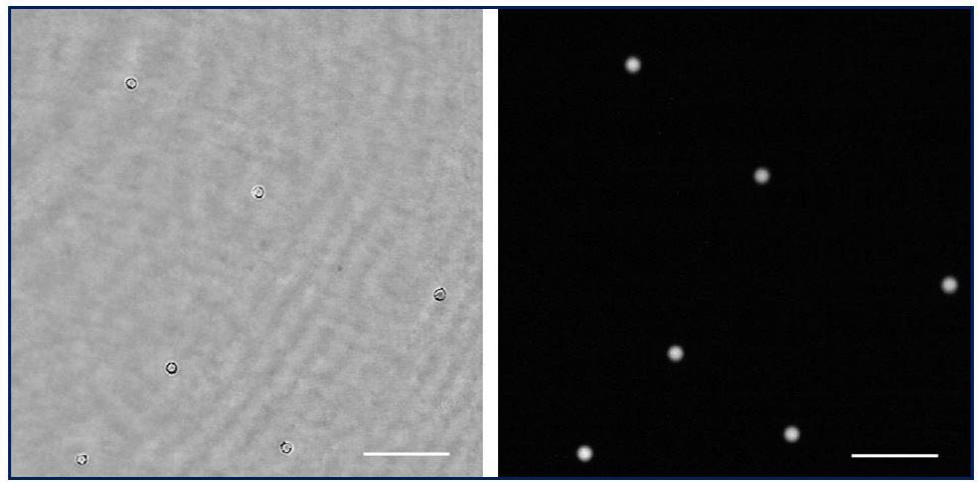

**Figure 9.** Simultaneous visualization of the phase retrieved by numeric propagation and fluorescence direct image of a low density fluorescent beads sample (Visualization 6). Scale bar is 50 µm length.

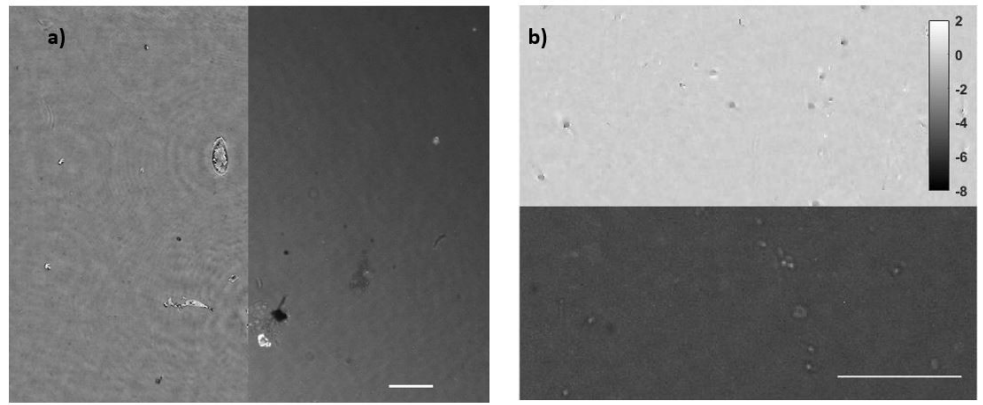

**Figure 10.** Movie videos of biological samples with the simultaneously combined image of holography and fluorescence. **(a)** Visualization of water suspended microorganism (Visualization 7) where left half FOV shows the retrieved phase distribution and right half FOV includes the fluorescence direct image. **(b)** Spermatozoa in free movement (Visualization 8) where bottom half FOV shows florescence and the upper half FOV includes phase distribution. White bars represent a length of 50 μm.

## 4. Discussion and conclusions

Through this manuscript we have proposed, calibrated and validated an extremely simple and cost-effective imaging platform integrating FI and QPI while using a single illumination light for both imaging modalities. The system has been optimized by the calibration of the optimum defocus distance which allows to obtain the best retrieved phase distribution from the in line holographic recording layout. Although calibration has been performed for a specific imaging system configuration, it is quite simple to apply similar procedures to variations of the system presented in this manuscript according to the target application. Thus, for instance, it is possible to select a different excitation wavelength depending on the fluorochrome to be used and to select the proper bandpass filter to separate both imaging modalities or to use a higher magnification and NA imaging lens to see smaller details in the samples. Whatever the case, we think the proposed concept defines a cost-effective and compact (it minimizes the number of optical elements) station which can be configurable depending on the type of samples to be imaged, fluorophores to be added and imaging requirements of the target application.

Worthy to notice is also the fact that the proposed concept can be perfectly implemented in regular/commercial bright field microscopes because all the needed modifications are implemented previous to the sample plane (inclusion of a coherent illumination source) and after the tube lens system (separation and digital recording of both imaging paths). Thus, it is possible to think in the external insertion of coherent illumination and a compact add-on module to be coupled at the exit port of the microscope embodiment for providing both simultaneous recordings. Since nowadays upgrading a regular microscope with other imaging modalities different from the ones provided by the microscope is a very appealing research field ([32]), this possibility will be explored in future works.

Regarding the layout, the introduction of the defocus distance (11 cm in our case) could be a size problem making a bit bulky the whole system. However, on one hand, it is the simplest way to achieve holographic imaging and, on the other hand, it is possible to reduce such defocus distance by including additional optics at the expenses of increasing a bit the price of the imaging platform. Also, from the retrieved quantitative phase images, we can observe some oscillations in the optical phase profile that comes from coherent noise and the influence of the twin image caused by the in-line Gabor holography scheme. These problems can lead into less accurate results but there are some solutions to be implemented. Coherent noise can be strongly mitigated by the use of partial coherent illumination [35], so that there are not so many interferences between the waves that are back and forth reflected. Furthermore, to increase the quality of the retrieved phase distributions and minimize the twin image effect, there are some different phase recovery algorithms that raise, for example, the use of several wavelengths during holographic reconstruction [36]. Whatever the case, we have computed the STD values at clear background areas in some of the holographic reconstructions as a metric related with the phase accuracy in the proposed holographic layout. The STD results are as follows: 0.14 rads (Fig. 2a.1), 0.13/0.17/0.18 rads (Figs. 4a - left to right), 0.13 rads (Fig. 5b.1) and 0.19 rads (Fig. 5e), yielding in a averaged STD value of 0.16 rads considering all images. This is a nice value considering the Gabor restriction defined in the holographic architecture. For instance, 0.16 rads are equal to 20 nm when considering height variation in the system (see Table 1) and, although holography can be much more accurate in general, this is good compromise between accuracy and technique complexity from a experimental point of view.

Anyway, note that our analysis considers thin samples. In the case of having volume information or thick samples, two constraints raise. First, thick/volume samples must still satisfy Gabor's condition since our holographic reconstruction is based on weak diffraction assumption. For this reason, density of the sample cannot be too high and occlusions (objects or particles projecting its own shadow one to each other) should not happen in order to preserve a reasonable Gabor behavior in holography. And second, thick/volume samples will provide more than a single sample section with useful information. Since the optimum defocus distance here derived is for a single plane (let's say the central plane of the thick sample), sections before and after that central plane will not be under optimum defocus distance recording conditions. Fortunately, the proposed system is not too much sensible to this fact because there is a range in defocus distances that seems to apply pretty well in holographic reconstruction ( 700 < Δz < 1450 µm according to Fig. 3), so the presented approach perfectly tolerates this fact. Nevertheless, this is only a constraint to be considered in the holographic imaging mode because fluorescent imaging does not allow 3D image over a thick/volume sample and only provides image of a single image plane section can be obtained (using a single recording). So in the limitation introduced by having dual mode imaging of the same section, the proposed approach is perfectly defined.

Moreover, since we are defining different imaging recording conditions on both imaging modes, the magnification provided by each mode is slightly different. FI provides a magnification according to the used infinity corrected microscope which is close to a 10X factor since we are using a tube lens with 180 mm focal length. However, QPI defines a Gabor in-line hologram which, in addition, introduces an intrinsic magnification factor as the geometrical projection of the object (in our case, the image provided by our microscope layout) over the camera plane. This geometric magnification ($M_G$) depends on the ratio between the distance from the illumination point source to the digital camera (d) and the distance between the illumination point source to the object (z) in the form of $M_G$ = d/z. In our case, the effective point source position comes from the image of the fiber optic end through the whole system and it is theoretically located at infinity because the tube lens is aimed to provide collimated illumination at the image space (blue beam at Fig. 1). Under these conditions, $M_G$ = 1. However, this position can be slightly altered due to experimental reasons, so that $M_G$ will be slightly different from 1 because of the definition of $\Delta z'$ = 11 cm. This is the reason why the FOVs provided by both imaging modalities are not exactly the same ones as can be appreciated from Figs. 8 (a.1) and (b.1).

In the case of the fluorescent imaging mode, we have used a fluorochrome type (fluorescein) which is commonly used in ophthalmology but not in biology or biomedicine. Because of this, maybe the results are not significant from a biological point of view but, in our view, they are good enough to show fluorescent effect and its simultaneous combination with holography in a single imaging platform. For sure that a specific fluorophore will have a higher adhesion to the cells structures showing more significant information as usually happen in biology. But there are a bunch of fluorophores with potential use in our presented imaging platform concept because excitation and emission are away enough to be separated with a long-pass filter, thus the proposed concept prevails. Here, the use of a generic fluorophore has the main objective of showing that it is possible to capture the simultaneous image of FI and QPI modes using only one illumination light.

Finally it is important to highlight that despite the fact that we have proposed a dual mode station mixing holography and fluorescence in optical microscopy (both imaging tech- niques working simultaneously), the user can decide to work only with one of them. In that case, the imaging platform will act as a regular microscope updated with coherence sensing capabilities, that is, as a holographic microscope restricted to a Gabor's domain. In that sense, in-line holography allows QPI at different sample sections. So, as proposed, the imag- ing platform is capable of simultaneously working with full 3D holographic reconstructions and single plane fluorescent imaging. However, we have in mind the development of novel methodologies in future works involving 3D fluorescent imaging aided by holographic information. Since fluorescence is an incoherent imaging technique, 3D reconstructions for thick samples usually need axial scanning of the input volume. This fact penalizes the temporal resolution of the method preventing the analysis of fast/dynamic events. We think that holography can help fluorescence to achieve digital 3D reconstructions from a single image plane recording. And with this manuscript, we are presenting the first steps towards this direction by introducing the common station for achieving simultaneous fluorescent and holographic imaging in a compact and cost-effective way.

In summary, we have reported on an experimental dual mode imaging platform the simultaneous FI and QPI analysis of static and living samples. The key concept is the use of a single illumination wavelength for providing both the excitation light in FI and the coherent beam for QPI. This characteristic confers the system with some advantages con- cerning compactness, simplicity and pricing of the complete imaging platform. For a wide variety of samples, we have provided, first, calibration and analysis of the independent imaging modes and, after that, validation of FI and QPI while working simultaneously in parallel (single shot imaging capability).

**Author Contributions:** All coauthors had contributed, with similar participation, to the following manuscript parts: conceptualization, methodology, numerical processing, validation, data analysis and data curation. Aditionally: Writing—original draft preparation, D.A. and V.M.; writing—review and editing, V.M. and J.G.; resources, J.G. and V.M.; supervision, J.G. and V.M.; project administration, J.G. and V.M.; funding acquisition, J.G. and V.M. All authors have read and agreed to the published version of the manuscript.

**Funding:** Grant PID2020-120056GB-C21 funded by MCIN/AEI/10.13039/501100011033.

**Data Availability Statement:** Not applicable.

**Acknowledgments:** D.A. acknowledges an internal scholarship at the University of Valencia (Ayudas para la colabroación en la investigación, Curso 21-22, Expediente Num. 1879967) to do this work .

**Conflicts of Interest:** The authors declare no conflict of interest.

### Abbreviations

The following abbreviations are used in this manuscript:
FI      Fluorescence imaging
QPI   Quantitative phase imaging
NA    Numerical aperture
FOV  Field of view
USAF United State Air Force
CAM  Camera

## References

1. Vogler, N.; Heuke, S.; Bocklitz, T.W.; Schmitt, M.; Popp, J. Multimodal imaging spectroscopy of tissue. *Annual Rev Anal Chem* **2015**, *8*, 359–387.
2. Jena, B.P.; Taatjes, D.J. *NanoCellBiology: Multimodal Imaging in Biology and Medicine*, 1st ed.; Pan Stanford Publishing: Penthouse Level, Suntec Tower 3, 8 Temasek Boulevard, Singapur, 2014.
3. Zhou, Q.; Chen, Z.; *Multimodality Imaging: For Intravascular Application*, 1st ed.; Springuer: Midtown Manhattan, New York City, United States, 2020.
4. Sen H.N.; Read, R.W.; *Multimodal Imaging in Uveitis*, 1st ed.; Springuer: Midtown Manhattan, New York City, United States, 2018.

5. Suresh, A.; Udendran, R.; Vimal, S. *Deep neural networks for multimodal imaging and biomedical applications*, 1st ed.; IGI Global: Pensilvania, United States, 2020.
6. Baz, A.E-.; Suri, J.S.*Big Data in Multimodal Medical Imaging*, 1st ed.; Chapman Hall: London, Unite Kingdom, 2020.
7. Mazumder, N.; Balla, N.K.; Zhuo, G.-Y.; Kistenev, Y.V.; Kumar, R.; Kao,F.-J.; Brasselet, S.; Nikolaev, V.V.; and Krivova, N.A. Label-Free Non-linear Multimodal Optical Microscopy—Basics, Development, and Applications. *Front. Phys.* **2019**, *7*, 170.
8. Yu, Z.G-.; Spandana, K.U., Sindhoora, K.M., Kistenev, Y.V., Kao, F.-J.; Nikolaev, V.V.; Zuhayri, H.; Krivova, A.N. and Mazumber, N. Label-free multimodal nonlinear optical microscopy for biomedical applications. *J. Appl. Phys.* **2021**, *129*, 214901.
9. Peres, C.; Nardin,C.; Yang,G. and Mammano, F. Commercially derived versatile optical architecture for two-photon STED, wavelength mixing and label-free microscopy. *Biomed. Opt. Express* **2022**, *13*, 1410-1429.
10. Rodríguez, A.D.; Clemente, P.; Tajahuerce, E.; Lancis, J. Dual-mode optical microscope based on single-pixel imaging. *Opt. Lasers. Eng.* **2016**, *82*, 87-94.
11. Picazo-Bueno, J.A.; Cojoc, D.; Iseppon, F.; Torre, V. and Micó, V. Single-shot, dual-mode, water-immersion microscopy platform for biological applications. *Appl. Opt.* **2018**, *57*, A242-A249.
12. Yeh, L.-H.; Chowdhury, S. Repina, N.A. and Waller, L. Speckle-structured illumination for 3D phase and fluorescence computational microscopy. *Opt. Express.* **2019**, *10*, 3635-3653.
13. Lee, Y.; Kim, B.; Choi, S. On-Chip Cell Staining and Counting Platform for the Rapid Detection of Blood Cells in Cerebrospinal Fluid. *Sensors* **2018**, *18(4)*, 1124.
14. Pavillon, N.; Fujita, K.; Smith, N.I; Multimodal label-free microscopy. *J. Innov. Opt. Health Sci.* **2014**, *7(5)*, 1330009.
15. Krafft, C.; Schmitt, M.; Schie, I.W.; Cialla, D.-M.; Matthaus, C.; Bocklitz, T. Label-Free Molecular Imaging of Biological Cells and Tissues by Linear and Nonlinear Raman Spectroscopic Approaches. *Angew. Chem. Int. Ed.* **2017**, *56(16)*, 4392–4430.
16. Ryu, J.; Kang, U. Song, J.W.; Kim, J. Kim, J.M.; Yoo, H. and Gweon, B. Multimodal microscopy for the simultaneous visualization of five different imaging modalities using a single light source. *Biomed. Opt. Express* **2021**, *12*, 5452-5469.
17. Park, Y.; Popescu, G.; Badizadegan, K.; Dasari, R.R. and Feld, M.S. Diffraction phase and fluorescence microscopy. *Opt. Express* **2006**, *14*, 8263-8268.
18. Yelleswarapu, C.S.; Tipping, M.; Kothapalli, S.-R.; Veraksa, A. and Rao, D.V.G.L.N. Common-path multimodal optical microscopy. *Opt. Lett.* **2009**, *34*, 1243-1245.
19. Quan, X.; Nitta, K.; Matoba, O.; Xia, P. and Awatsuji, Y. Phase and fluorescence imaging by combination of digital holographic microscopy and fluorescence microscopy. *Opt. Lett.* **2015**, *22(2)*, 349–353.
20. Chowdhury, S.; Eldridge, W.J.; Wax, A. and Izatt, J.A. Spatial frequency-domain multiplexed microscopy for simultaneous, single-camera, one-shot, fluorescent, and quantitative-phase imaging. *Opt. Lett.* **2015**, *40*, 4839-4842.
21. Chowdhury, S.; Eldridge, W.J.; Wax, A. and Izatt, J.A. Structured illumination multimodal 3D-resolved quantitative phase and fluorescence sub-diffraction microscopy. *Biomed. Opt. Express* **2017**, *8*, 2496-2518.
22. Shen, Z.; He, Y. Zhang, G. He, Q. Li, D. and Ji, Y. Dual-wavelength digital holographic phase and fluorescence microscopy for an optical thickness encoded suspension array. *Opt. Lett.* **2018**, *43*, 739-742.
23. Dubey, V.; Ahmad, A.; Singh, R.; Wolfson, D.L.; Basnet, P.; Acharya, G.; Mehta, D.S. and Ahluwalia,B.S. Multi-modal chip-based fluorescence and quantitative phase microscopy for studying inflammation in macrophages. *Opt. Express* **2018**, *26*, 19864-19876.
24. Nygate, Y.N.; Singh, G.; Barnea, I and Shaked, N.T. Simultaneous off-axis multiplexed holography and regular fluorescence microscopy of biological cells. *Opt. Lett.* **2018**, *43*, 2587-2590.
25. Kernier, I.D.; Cherif,A.A-.; Rongeat, N.; Cioni, O.; Morales, S.; Savatier, J.; Monneret, S. and Blandin, P. Large field-of-view phase and fluorescence mesoscope with microscopic resolution.*J. Biomed. Opt.* **2019**, *24*, 036501.
26. Mandula, O.; Kleman, J-.P.; Lacroix, F.; Allier, C. Fiole, D.; Hervé, L.; Blandin, P.; Kraemer, D.C. and Morales, S. Phase and fluorescence imaging with a surprisingly simple microscope based on chromatic aberration.*Opt. Express* **2020**, *28*, 2079-2090.
27. Kumar, M.; Quan, X.; Awatsuji, Y.; Tamada, Y. and Matoba, O. Digital holographic multimodal cross-sectional fluorescence and quantitative phase imaging system. *Sci. Rep.* **2020**, *10*, 1–13.
28. Wen, K.; Gao, Z.; Fang, X.; Liu, M.; Zheng, J.; Ma, M.; Zalevsky, Z. and Gao, P. Structured illumination microscopy with partially coherent illumination for phase and fluorescent imaging. *Opt. Express* **2021**, *29(21)*, 33679-33693.
29. Quan, X.; Kumar, M.; Rajput, S.K.; Tamada, Y.; Awatsuji, Y. and Matoba, O. Multimodal microscopy: Fast acquisition of quantitative phase and fluorescence imaging in 3D space. *IEEE J. Sel. Topics Quantum Electron* **2021**, *27(4)*, 6800911.
30. Liu, P.Y. et al. Cell refractive index for cell biology and disease diagnosis: past, present and future. *Lab Chip* **2016**, *16*, 634–644.
31. Lichtman J.W. and Conchello, J.-A. Fluorescence microscopy. *Nat. Methods* **2005**, *2*, 910–919.
32. Micó, V.; Trindade, T. and Picazo-Bueno, J.A. Phase imaging microscopy under the Gabor regime in a minimally modified regular bright-field microscope. *Opt. Express* **2021**, *29(26)*, 42738-42750.
33. Trusiak, M.; Cywińska, M.; Micó, V.; Picazo-Bueno, J.Á.; Zuo, C.; Zdańkowski, P.; Patorski, K. Variational Hilbert quantitative phase imaging. *Sci. Rep.* **2020**, *10(1)*, 13955.
34. Zhai, J.; Shi, R.; Kong, L. Improving signal-to-background ratio by orders of magnitude in high-speed volumetric imaging in vivo by robust Fourier light field microscopy. *Photon. Res.* **2022**, *10, 1255-1263*.
35. *Perucho, B.; Micó, V. Wavefront holoscopy: application of digital in-line holography for the inspection of engraved marks in progressive addition lenses* J. Biomed. Opt. **2014**, *19(1), 016017.*
36. *Sanz, M.; Picazo-Bueno, J.A.; García, J.; Micó, V. Dual mode holographic microscopy imaging platform* Lab Chip **2018**, 18(7), *1105-1112.*